\documentclass[11pt,a4paper]{article}

\usepackage[utf8]{inputenc}
\usepackage[T1]{fontenc}
\usepackage{lmodern}
\usepackage{geometry}
\usepackage{setspace}
\usepackage{hyperref}
\usepackage{graphicx}
\usepackage{longtable}
\usepackage{caption}
\usepackage{enumitem}
\usepackage{titlesec}
\usepackage{csquotes}
\usepackage{cleveref}
\usepackage{booktabs}   
\usepackage{array}      
\usepackage{tabularx}   
\usepackage{adjustbox}  
\usepackage{authblk}

\usepackage[numbers]{natbib}

\hypersetup{colorlinks=true, linkcolor=blue, citecolor=blue, urlcolor=blue}

\title{How Do Ports Organise Innovation? Linking Port Governance, Ownership, and Living Labs}
\author[1]{Sonia Yeh\thanks{Corresponding author: \texttt{sonia.yeh@chalmers.se}}}
\author[2]{Christopher Dirzka}
\author[1]{Aleksandr Kondratenko}
\author[1]{Frans Libertson}
\author[3]{Benedicte Madon}

\affil[1]{Chalmers University of Technology, Gothenburg, Sweden}
\affil[2]{Copenhagen Business School, Copenhagen, Denmark}
\affil[3]{Universidad de Sevilla, Seville, Spain}

\begin{document}

\begin{titlepage}
  \maketitle
  \vfill

  \noindent\textbf{Acknowledgements:} 
We thank the European Network of Living Labs (ENoLL) facilitation team, including Elina Makousiari and Mar Ylla, for designing and moderating the capacity-building activities and for synthesis support that ensured methodological consistency across sites. We are grateful to colleagues at the Port of Aalborg (PoA), especially Lars J{\o}ker, Palle Skyum, and Maria Risager Nielsen, and at the Port of Trelleborg (PoT), including Jennie Folkunger, for their collaboration and openness throughout workshops and follow-up sessions. We also acknowledge the many stakeholders from the public sector, industry and terminal operators, academia, and civil society who contributed their time and insights.

This work received financial support from the Clean Energy Transition Partnership (CETPartnership, Grant 101069750 under HORIZON-CL5-2021-D3-01-04), the Swedish Energy Agency, Danish Innovationfonden, and the Ministerio de Ciencia, Innovaci{\'o}n y Universidades. The views expressed are those of the authors and do not necessarily reflect those of the funding organisations. Any remaining errors are our own.


\end{titlepage}

\setcounter{page}{1} 

\makeatletter
\begin{center}
  {\LARGE\bfseries \@title \par}
\end{center}
\makeatother

\begin{abstract}
Ports are pivotal to decarbonisation and resilience. Existing port studies document sustainability and digital pilots, but rarely examine how ownership and decision rights shape the process and outcomes. In parallel, Living Lab (LL) scholarship provides strong conceptual foundations, but offers limited sector-grounded explanation of \emph{LL--governance fit} in ports. This paper addresses these gaps by developing and applying a governance--LL fit framework that links port governance and ownership to four LL pillars: co-creation, real-life setting, iterative learning, and institutional embedding (multi-level decision-making). We apply the framework in a comparative case study of two analytically contrasting ports. To ground the analysis in port-defined priorities, we use two case anchors: an \emph{Energy Community} pilot in Aalborg and a \emph{Green Coordinator} function in Trelleborg. We interpret these anchors through the LL three-layer logic (macro--meso--micro), distinguishing the stable constellation of actors and mandates (macro), the governance of specific projects (meso), and the methods used to generate and test solutions (micro). Findings show that Landlord governance offers contract- and procurement-based ``landing zones'' (concessions/leases and tender clauses) that can codify LL outputs and support scaling across tenants and infrastructures, whereas Tool/Public Service governance embeds learning mainly through SOPs, procurement specifications, and municipal coordination, enabling internal operational gains but limiting external codification without bespoke agreements. Despite these differences, both ports share core LL building blocks and face similar needs for clear role definition, sustained stakeholder engagement, and timely alignment with decision windows. Overall, LL effectiveness emerges as governance-contingent rather than governance-neutral. More broadly, the findings suggest that living-lab effectiveness in large infrastructure organisations is contingent on governance—specifically, where decision rights sit and which instruments exist to embed learning into routine practice.
\end{abstract}

\section{Introduction}

Nearshore infrastructure developments, such as maritime ports and inland ports, are critical nodes in global transport and trade networks that enable the movement of goods, energy, and people while underpinning regional and national economies. They are also increasingly recognised as key arenas for addressing sustainability challenges, including the decarbonisation of transport, the integration of renewable energy, and the creation of more resilient supply chains \cite{Monios2024}. Yet, ports display significant variation in their capacity to pursue sustainability transitions \cite{Monios2024, LamNotteboom2014, Munim2020}. A key factor underlying this divergence is governance, which is inherently linked to infrastructure ownership, an element that is central but frequently implicit. Ownership determines who exercises property and decision rights over land, berths, energy systems, digital platforms, and grid connections. These governance arrangements are decisive in shaping how ports manage shifts in energy and digital infrastructure.

Port governance defines the distribution of decision-making authority, the scope of public and private responsibilities, and the mechanisms through which stakeholders interact \cite{Sornn-Friese2024}. These institutional arrangements shape not only operational efficiency and investment decisions but also the capacity for innovation and the integration of new technologies and practices. In particular, they determine the ease or difficulty with which ports can foster multi-stakeholder collaboration and embed experimental approaches into ongoing operations.

In this context, \textit{Living Labs} (LLs) have emerged as a prominent innovation model for organising multi-actor experimentation in complex socio-technical settings. LLs bring together diverse stakeholders in real-life environments to co-create, test, and refine solutions \cite{ENoLL2025}. However, LL effectiveness should not be assumed to be governance-neutral: ports differ in their mandates, decision rights, contracting arrangements, and ownership structures, which may systematically condition whether LL outputs can be translated into durable changes in rules, investments, and operations.

This article examines how port governance models---Public Service, Tool, Landlord, and Private---shape the feasibility and design of LL approaches, using evidence from structured workshops and stakeholder engagement activities at the Port of Aalborg (PoA), Denmark, and the Port of Trelleborg (PoT), Sweden. We develop a framework linking governance characteristics to LL core pillars, then apply it to two illustrative cases: PoA, operating under a Landlord model, and PoT, operating closer to a Tool/Public Service model. The analysis highlights how infrastructure ownership and control are not merely background conditions but actively shape the governance--LL fit and influence whether experimental initiatives remain project outputs or scale into more systemic change.

Although developed in port settings, the framework is applicable to other infrastructure-intensive organisations where ownership and decision rights shape experimentation and adoption---for example airports, rail terminals, logistics hubs, and science parks---particularly where Tool/Public Service-like governance constrains external codification and elevates the role of internal procedures and inter-agency coordination.

\subsection{Living Labs as an innovation-governance model}
\label{subsec:ll_innovation_model}

The European Network of Living Labs (ENoLL) defines LLs as ``open innovation ecosystems in real-life settings, using iterative feedback processes throughout the innovation lifecycle to create sustainable impact'' \cite{ENoLL2025Origins}. This framing emphasises that LLs are not merely project formats but ecosystem arrangements intended to connect research, industry, policy, and society in collaborative experimentation.

Recent scholarship has deepened the understanding of LLs as instruments of innovation governance rather than only co-creation venues. In this view, LLs function as structured, experimental ``test beds'' that convene state and non-state actors to trial solutions under real-world constraints while keeping outputs provisional and adaptable \cite{Engels2019}. Complementing this governance lens, LLs are commonly situated in the \emph{Quadruple Helix} model (public sector, industry, academia, and civil society/users), which highlights that innovation depends on coordination across institutional spheres beyond the firm \cite{CavalliniEtAl2016QuadrupleHelix}.

Building on the Quadruple Helix view of multi-sphere participation, governance in LLs is often described across three layers (macro--meso--micro). The macro level concerns the LL constellation (partners, legitimacy, and positioning in the wider ecosystem); the meso level concerns the governance of specific LL projects; and the micro level concerns methodological steps such as co-creation, real-life testing, and evaluation \cite{Schuurman2015PhD}. This layered view is particularly useful for ports because it helps distinguish (i) ecosystem-level participation and legitimacy questions (macro), (ii) project-level steering and commitments (meso), and (iii) the design and execution of trials (micro).

Finally, building on ecosystem perspectives, LLs can be conceptualised as \emph{orchestrators} within regional innovation systems, shaping alignment, collaboration, and proof-of-concept pathways that connect experimentation to wider system change \cite{FauthDeMoortelSchuurman2024Orchestrators}. This orchestrator framing is especially salient in ports, where decision rights and ownership arrangements can condition whether learning at the micro level can be translated into binding instruments at meso and macro levels (e.g.\ procurement specifications, operating procedures, and concession or lease clauses).

\subsection{Living Labs in port contexts}
\label{subsec:ports_ll_literature}

Ports increasingly appear as real-life settings for LL-type experimentation, particularly in (i) climate resilience and disruption preparedness, where LLs are used to convene stakeholders and co-create adaptation solutions in operational environments \citep{PolydoropoulouEtAl2025PortsLLResilience}, (ii) sustainability transitions (including proposals for trans-local or multi-port LL configurations), often emphasising learning and diffusion beyond single sites \citep{GerlitzMeyerHenesey2024SMSPsLivingLabs}, and (iii) digital transformation and logistics innovation, where LL cases are mobilised to link technical experimentation with policy learning and stakeholder coordination \citep{CatanaMerloPerboli2025GoverningDigitalTransformationPorts}.
Alongside this academic work, practice-led port LLs have been established by port authorities to support experimentation under real operating conditions (e.g.\ Singapore and Halifax), underscoring practical momentum but typically offering limited analytical leverage for explaining why some experiments translate into durable changes while others remain project outputs \citep{MPA2026LivingLab, WPSP2023PIERLivingLab}.

Overall, the literature remains comparatively thin on an explicit, sector-grounded account of \emph{LL--governance fit} in ports---that is, how port governance and ownership structures shape (a) meaningful participation, (b) the feasible scope of system-level experimentation, and (c) the institutional embedding of evidence into multi-level decision-making instruments (e.g.\ procurement specifications, operating procedures, and concession or lease arrangements). This paper addresses that gap by analysing LL--governance fit from a ports perspective, linking LL pillars to port governance and ownership models as an explanatory lens for variation in outcomes.

\subsection{Port governance models and the research gap on LL--governance fit}
\label{subsec:port_governance_gap}

\begin{figure}[h!]
\centering
\includegraphics[width=1\linewidth]{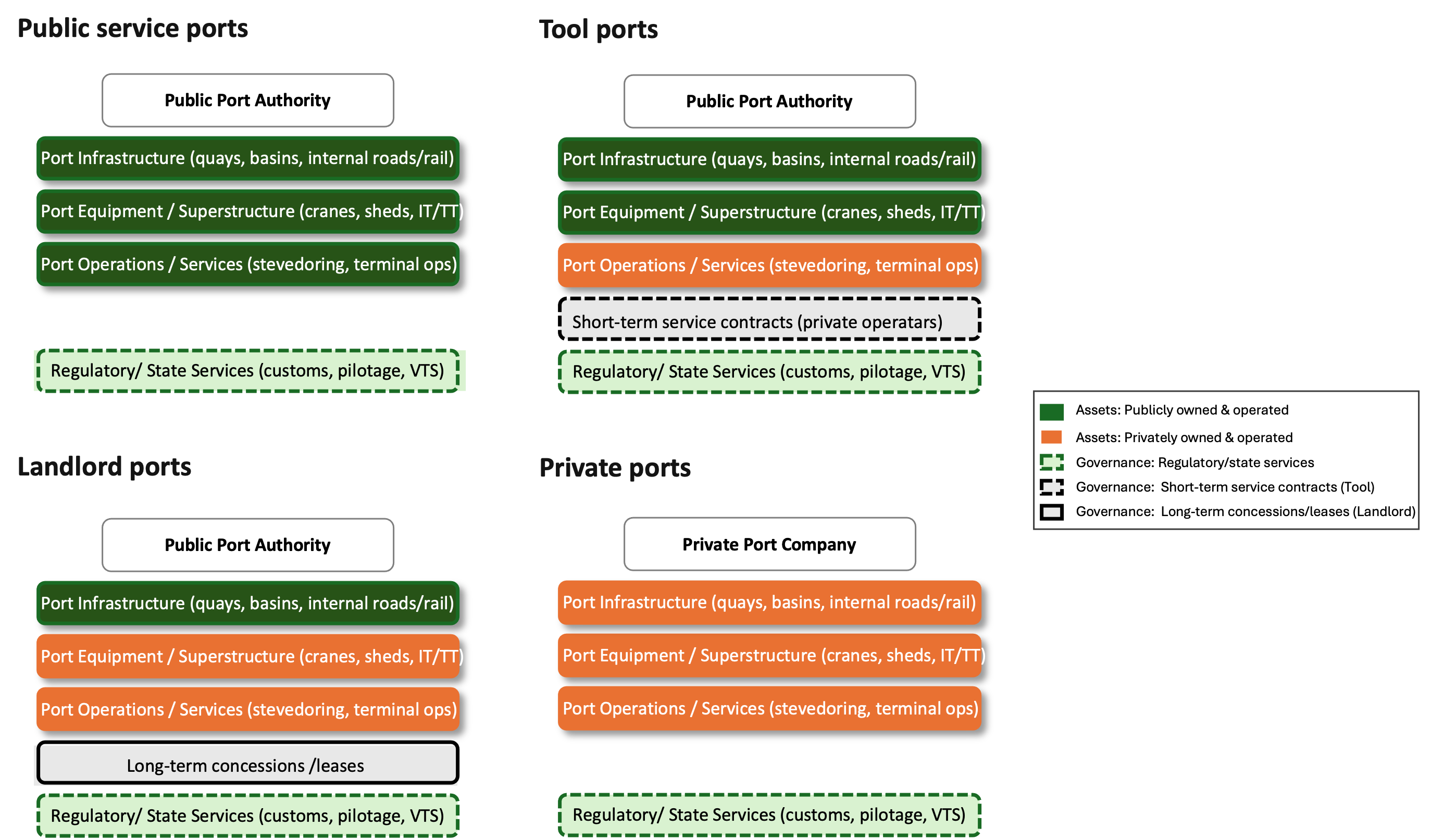}
\caption{Organisational scheme of port governance models.
Dark green = public; orange = private across infrastructure, equipment/superstructure, and operations. Dashed green box = regulatory/state services (e.g.\ customs, pilotage, Vessel Traffic Service (VTS)). Banners show contract type: dashed = short-term (Tool), solid = long-term (Landlord).}
\label{fig:landlord}
\end{figure}

Port governance refers to the institutional arrangements that define infrastructure ownership, allocate operational responsibilities, and shape decision-making processes (Figure~\ref{fig:landlord}). Infrastructure ownership---whether public, private, or hybrid---determines who has control over assets like berths, terminals, substations, data infrastructure, and energy systems. These arrangements influence strategic priorities, investment patterns, and the scope for stakeholder collaboration. A widely used classification, developed by the World Bank \cite{WorldBank2007} and refined by Notteboom, Pallis, and Rodrigue \cite{Notteboom2021}, identifies four main governance models, differentiated primarily by the division of public and private responsibilities (Table~\ref{tab:models}).

\begin{table}[h]
\centering
\small
\caption{Port governance models and examples (based on \cite{Notteboom2021}).}
\label{tab:models}
\begin{tabularx}{\textwidth}{@{} l X X @{}}
\toprule
\textbf{Governance Model} & \textbf{Description} & \textbf{Example Ports} \\
\midrule
Public Service & Fully public ownership; authority controls operations, maintenance, and investment. Rare today---mostly historical or small national ports. & Port of Shanghai \\
Tool & Authority owns infrastructure and equipment; private operators deliver services via short contracts. & Port of Trelleborg; Port of Kristiansand \\
Landlord & Authority plans and owns land/infrastructure; private firms operate terminals under lease/concession. & Port of Rotterdam; Port of Gothenburg; Port of Hamburg; Port of Singapore \\
Private & Private ownership of infrastructure and operations; public sector limited to regulation. & Common in UK ports; Port of Brofjorden (Preemraff Lysekil) \\
\bottomrule
\end{tabularx}
\end{table}

\textbf{Public Service Ports} Entirely publicly owned and operated. The port authority manages infrastructure, superstructure, and operations. The model was naturally created by state or colonial powers to control and develop ports and was dominant from the 19th century to the mid-20th century \cite{WorldBank2007}. Although the model is now rare in major commercial ports, some of the leading ports of the world are still using it. The most prominent example is the Port of Shanghai, the largest container port in the world.

\textbf{Tool Ports}
The port authority owns the infrastructure and equipment, while private operators provide specific services (e.g., cargo handling) under short-term contracts. This model makes the first step in the transition from fully public port governance towards better performance by incorporating the private entities into the process. Examples include the Port of Trelleborg---the largest Ro-Ro port in Scandinavia---and the Port of Kristiansand in Norway.

\textbf{Landlord Ports}
The port authority owns and develops land and infrastructure, and leases terminals to private operators under long-term concessions. This has become the most prevalent and successful model globally. Often, the Landlord model provides the most efficient combination of public and private governance. Prominent examples are the Port of Singapore and the Port of Rotterdam.

\textbf{Private Ports}
Privately owned and operated. The public sector plays a limited role, largely confined to regulatory oversight and some public services such as customs and pilotage. A Private port supports the business interests of its owner. A typical example is the Port of Brofjorden---the largest oil port in Sweden---which belongs to the Preemraff Lysekil oil refinery.

\textbf{Hybrid/JV models}
A hybrid/JV port model blends public and private roles through contract forms such as concessions. Typically, the public sector retains land and core infrastructure rights, while private partners finance, build, and operate superstructure (e.g., terminals, cranes, energy assets) under long-term, performance-based agreements. Governance is shared with revenues and risk allocation tailored to each asset (tariff, demand, construction, and O\&M risk). This structure can mobilise capital, speed innovation (e.g., energy hubs/DSO joint ventures), and align incentives, while keeping public oversight on access, safety, and externalities. Its main challenges are coordination costs, interface risks between asset owners/operators, and the need for robust regulation and transparency to prevent monopoly rents and mission drift.

These governance configurations imply systematic differences in (i) who can authorise and fund experiments, (ii) how easily external stakeholders can be engaged, and (iii) which instruments exist to embed outcomes (e.g.\ through procurement, operating rules, or concession/lease structures). However, despite growing attention to LLs in port contexts (Section~\ref{subsec:ports_ll_literature}), existing work rarely examines \emph{how} port governance and ownership models condition LL design choices and the institutional embedding of LL results. We therefore develop and apply a governance--LL fit framework to compare two ports operating under different dominant governance models (PoA: Landlord; PoT: Tool/Public Service), with the aim of identifying where misalignments arise and what this implies for LL implementation and institutionalisation in ports.

\subsection{Research aims and scope}

This study has two aims: an analytical aim and an empirical aim. Analytically, we develop and operationalise a governance--LL fit framework that specifies how ownership and decision rights condition LL design, stakeholder intensity, and institutional embedding in ports. Empirically, we apply the framework to two contrasting port governance contexts (PoA; PoT) and examine where governance facilitates or impedes LL institutionalisation in energy and digital transition initiatives.

We address three research questions (RQs): 

\indent{RQ1 (Conditions):} How do governance and ownership arrangements (e.g.\ Public Service/Tool and Landlord/Private) shape stakeholder dynamics, decision authority, innovation scope, and mechanisms for embedding LL outcomes? \\[0.3em]
\indent{RQ2 (Mechanisms):} Through which contractual and procedural mechanisms (e.g.\ service contracts, concessions, procurement, internal procedures) are LL outputs translated into durable organisational change? \\[0.3em]
\indent{RQ3 (Design):} How should LL design (pillars, sequencing, artefacts) adapt to governance conditions to maximise the likelihood of institutional embedding?

We focus on PoA and PoT as analytically contrasting cases engaged in energy-transition initiatives under different dominant governance configurations (PoA: Landlord; PoT: Tool/Public Service). The analysis centres on early-stage institutionalisation dynamics---how ownership and decision rights shape stakeholder mobilisation, experimentation, and translation of evidence into decisions---rather than long-run performance evaluation. 

\section{Methods}
\label{sec:methods}

\subsection{Study design and cases}
We use a comparative case study design to examine how governance conditions shape the design and institutionalisation of LLs in ports. Two analytically contrasting cases were selected: Port of Aalborg (PoA), operating under a Landlord governance model, and Port of Trelleborg (PoT), operating closer to a Tool/Public Service configuration. The typology follows World Bank/Notteboom--Pallis--Rodrigue conventions \cite{WorldBank2007,Notteboom2021} to ensure comparability with established port governance literature. We treat each governance type as a dominant configuration; specific operational arrangements may temporarily resemble features of other models.

The unit of analysis is governance--LL fit within each port, namely how ownership and decision rights condition LL pillars (co-creation, real-life setting, iterative learning, and institutional embedding) and the mechanisms by which LL outputs are embedded into organisational decision instruments.

\subsection{Data}
Our empirical corpus integrates consortium onboarding, port-specific capacity-building workshops, and stakeholder-engagement instruments into a single, traceable dataset. A project-wide online ``Living Lab basics'' webinar on 11 March 2025 (120 minutes; $n=17$) established a common conceptual baseline (principles, terminology, expectations). Slides and notes informed subsequent port work.

A full-day, in-person workshop on 6 May 2025 was held at PoA followed by a focused online session on 30 May 2025 ($n=10$ across PoA, project partners, and facilitators). Outputs include completed canvases on purpose and mission/scope; host organisation and context; context mapping and SWOT; stakeholder identification, value-creation mapping, and initial problem statements. Paper artefacts were digitised to a shared collaborative board; the online session completed deferred items and consolidated actions.

A three-hour online workshop was held on 12 June 2025 at PoT, combining short primers with structured hands-on exercises, prioritising transition purpose, stakeholder mapping with prioritisation (five critical stakeholder groups identified), a stakeholder value matrix, and a streamlined SWOT. Members of the coordination team and peers from PoA joined portions for cross-case methodological consistency.

Two standardised instruments support follow-on data collection and member checks: (i) a value-proposition reflection sheet paired with an interview invitation template (to tailor outreach and clarify roles and benefits), and (ii) a semi-structured interview guide structured around role, collaboration experience, barriers and enablers, and conditions for future engagement. The dataset comprises webinar materials; two PoA workshops with digitised canvases (purpose, host context, mission/scope, SWOT, stakeholder mapping, value, problem statements); one PoT workshop with purpose, timeline, stakeholder mapping and prioritisation, value matrix, and SWOT; and the interview instruments and invitations. All artefacts are dated, attributed by session, and archived in shared boards and concise facilitator reports, enabling a clear chain of evidence from raw materials to assessment.


\subsection{Operationalisation}
The four dimensions were derived deductively by mapping core LL pillars onto governance mechanisms that determine participation, decision authority, feasible experimentation scope, and institutional embedding pathways in ports. Governance--LL fit is assessed along four dimensions, each paired to the LL pillars:

\begin{enumerate}
  \item Stakeholder dynamics (breadth and intensity across the Quadruple Helix, role clarity, conflict resolution) paired with co-creation.
  \item Decision authority and embedding (location of decision rights; availability of landing zones such as concessions, procurement, SOPs; timing windows) paired with institutional embedding (multi-level decision-making).
  \item Innovation scope (balance of tenant versus system-level pilots; energy--digital coupling; operational scale; external linkages) paired with real-life setting (system-level pilots require coordination across multiple port actors and/or infrastructures---e.g.\ port--utility--tenant arrangements, shared metering/data interfaces, or cross-tenant interoperability---rather than implementation by a single tenant or a single port unit).
  \item Institutional uptake mechanisms (contracts, concessions and leases, procurement clauses, SOPs and MOUs; specification quality; monitoring and enforcement; replicability) paired with iterative learning, with attention to how learning is codified into formalised instruments.
\end{enumerate}

Each dimension is scored on a 0--2 scale (0 = low/absent; 1 = moderate/mixed; 2 = high/strong) using explicit indicators. Short textual justifications tie scores to dated artefacts. The complete indicator set and scoring rubrics are provided in Appendix \ref{app:codebook}. 
Two researchers independently coded PoA and PoT using a structured template aligned with the four dimensions and LL pillars. Researcher~1 participated in all workshops, while Researcher~2 conducted offline coding using the workshop transcripts, digitised artefacts, and facilitator reports. We note that this difference in proximity to the live sessions may contribute to minor interpretive differences in how responses were read. Cross-case scores and evidence anchors are summarised in Appendix~\ref{app:matrix}.

Member checks were conducted by circulating one-page factual summaries per port to verify roles, processes, and constraints. All coding decisions maintain traceability to source materials. Cross-case comparison juxtaposes PoA and PoT scores in a cross-case matrix to identify systematic contrasts and design implications. Participation in workshops and interviews was voluntary. No sensitive operational data are reported without prior consent. Artefacts analysed were produced within project activities and are stored in controlled collaborative repositories. Identifiable statements are anonymised unless explicit permission was granted.

\section{Results}
\label{sec:results}

\subsection{Port of Aalborg (Landlord)}
PoA exhibits broad engagement across the Quadruple Helix with regular touchpoints among municipal entities, port companies and the port authority, terminal tenants, utilities, and research partners. Roles and responsibilities were articulated and iteratively refined through the capacity-building sequence, supporting common problem framing and sustained mobilisation around energy-system use cases. This configuration underpins strong co-creation and cumulative iterative learning, consistent with the LL emphasis on multi-actor orchestration in real-life environments.

The Landlord model provides multiple, well-specified entry points for embedding LL outputs, particularly via concession/lease structures and port-level procurement. Here, \emph{landing zones} refer to governance instruments such as concession/lease annexes, tender requirements, KPI and reporting clauses (e.g., energy performance, metering, data-sharing), and renewal windows where pilot evidence can be translated into enforceable obligations. Decision pathways are comparatively transparent and time-bound (e.g., concession renewal windows and capital budgeting cycles), enabling pilot evidence to flow into formal decision instruments and thereby strengthening institutional embedding (multi-level decision-making).

To make the results tangible at the use-case level, PoA has expressed interest in developing an \emph{Energy Community} as an early pilot. In LL terms, this sits at the \emph{meso} layer (project/pilot level), while the workshop activities and tools used to shape it (stakeholder mapping, mission setting, problem statements, etc.) sit at the \emph{micro} layer. The strategic and operational framing that stabilises how such pilots are selected, governed, and scaled corresponds to the \emph{macro} layer of the 3-layer LL model.

Overall, PoA’s pilot portfolio spans tenant-level interventions (e.g., electrified operations, data-sharing prototypes) and system-level initiatives (e.g., energy hubs, grid-service participation), with explicit coupling to digitalisation (data platforms, metering, interoperability). Trials occur in the operational environment at meaningful scale, aligning with the real-life setting pillar and facilitating progression from iterative learning to formalisation through contract-grade instruments.

\subsection{Port of Trelleborg (Tool/Public Service)}
PoT’s engagement is more targeted and formalised, anchored in the authority’s operational remit and statutory partnerships. ENoLL’s capacity-building pathway for PoT was explicitly more tailored than for PoA, reflecting differing starting points and internal capacity while still drawing on LL principles as a reference frame. Co-creation is present but benefits from deliberate facilitation to maintain cadence across cycles, especially when external actors’ participation depends on immediate project relevance and available resources.

Compared with a Landlord context, PoT affords fewer contract-grade landing zones for external institutionalisation. Centralised authority can support decisive internal pilots, but institutional embedding (multi-level decision-making) relies mainly on internal procedures (SOPs), procurement specifications, and municipal coordination rather than concession/lease mechanisms. In other words, institutional uptake is feasible for internal operations; external scaling typically requires bespoke arrangements (e.g., inter-organisational protocols, joint workplans, or MOUs where applicable).

PoT has indicated interest in using a \emph{Green Coordinator} function as a concrete case anchor. Conceptually, this differs from a single technical pilot: it is an organisational coordination mechanism intended to (i) stabilise ownership of sustainability initiatives, (ii) create continuity across projects, and (iii) provide an internal landing zone for translating LL learning into routines, procurement language, and cross-actor coordination. Within the 3-layer LL model, this is best understood as strengthening the \emph{macro} (operational governance capacity) while enabling more coherent \emph{meso}-level projects and repeatable \emph{micro}-level engagement activities.

Pilots in PoT concentrate on operational improvements within the authority’s control (e.g., traffic flow, shore power scheduling), with selective system-level links to DSOs and municipal energy planning. The real-life setting is strong for internal pilots, while system-level reach remains more constrained by governance and by the limited availability of external codification instruments. Iterative learning does translate into formalisation, but the scope of codification is narrower than in a Landlord setting.

\subsection{Cross-case patterns, case anchors, and common elements}
\label{sec:crosscase}

To keep the Results section tightly anchored in these empirical materials, we separate (i) \emph{evidence-anchored cross-case patterns} from (ii) \emph{design implications} that are developed analytically from those patterns. Accordingly, we retain the empirical cross-case pattern table (Table~\ref{tab:patterns}) and complement it with a case-to-three-layer mapping (Table~\ref{tab:layers}), which situates the two port-defined case anchors within the macro--meso--micro LL logic and the associated governance landing zones.

Across both ports, several common elements emerge from the workshop artefacts and interview materials. First, both LLs rely on a shared set of micro-level engagement tools---workshops, stakeholder identification and prioritisation, SWOT, and value mapping---to surface priorities, clarify roles, and translate broad transition ambitions into actionable problem statements. These tools were provided through the ENoLL capacity-building package, but they are also widely used participatory methods; their analytical value here lies in what they produce and how those outputs can (or cannot) be carried forward into governance-specific landing zones. Second, both ports faced a recurring need for clearer role definition and sustained reactivation of stakeholders over time, particularly to maintain participation beyond the initial framing phase. Third, both cases indicate that civil-society/user participation is not automatic: even with an explicit Quadruple Helix framing, engagement beyond public and industry actors requires deliberate design choices and outreach.

\begin{table}[h!]
\centering
\small
\caption{Cross-case patterns by governance--LL fit dimension (synthesis of coded results).}
\label{tab:patterns}
\begin{tabular}{p{3.2cm} p{5.8cm} p{5.8cm}}
\toprule
Dimension & Landlord context (PoA) & Tool/Public Service context (PoT) \\
\midrule
Stakeholder dynamics &
Broad, continuous engagement across the Quadruple Helix; role clarity supports parallel experimentation and portfolio logic. &
Targeted, formalised engagement; strong focus but cadence requires active facilitation to sustain iterative cycles. \\
Decision authority and embedding &
Multiple contractual and procedural landing zones (concessions, leases, procurement); clearer timing windows for codifying results. &
Fewer external landing zones; embedding channels are primarily internal procedures, procurement, and municipal coordination. \\
Innovation scope &
Balanced portfolio spanning tenant pilots and system-level initiatives; stronger energy--digital coupling and operational scale. &
Operational pilots in the authority’s remit with selective system linkages; system-level reach developing and more contingent. \\
Institutional uptake mechanisms &
Contract-grade instruments (concession/lease annexes, KPI clauses) and port-wide procedures enable diffusion and enforceability. &
Uptake via SOPs, procurement specifications, and coordination instruments; codification scope narrower and often bespoke. \\
\bottomrule
\end{tabular}
\end{table}

\begin{table}[h!]
\centering
\small
\caption{Grounding the case anchors in the LL three-layer model (macro--meso--micro) and governance ``landing zones''.}
\label{tab:layers}
\begin{adjustbox}{max width=\textwidth}
\begin{tabular}{p{3.2cm} p{3.3cm} p{3.7cm} p{3.7cm} p{3.9cm}}
\toprule
Port (governance) & Typical governance landing zones & Macro layer (stable constellation) & Meso layer (projects/pilots) & Micro layer (methods/tools) \\
\midrule
PoA (Landlord) &
Concession/lease annexes; KPI and reporting clauses; tender requirements; renewal windows. &
Port-based LL constellation anchored by port companies and the port authority with municipal, tenant, utility, and research participation; strategic and operational framing stabilises selection and scaling logic. &
\textbf{Energy Community} prioritised as an early pilot; adaptable project-level configuration to specific stakeholders and goals. &
Workshops, stakeholder mapping, mission/scope canvases, and problem statements. \\
\midrule
PoT (Tool/Public Service) &
SOP updates; procurement specifications; municipal/inter-agency coordination instruments; bespoke agreements where needed. &
Port authority anchored in operational remit and statutory and municipal linkages; tailored LL pathway reflecting organisational starting point. &
\textbf{Green Coordinator} proposed as an organising case anchor: an internal coordination function intended to stabilise sustainability work across initiatives and partners (enabling coherent project selection and continuity). &
Workshops, stakeholder identification and prioritisation, value matrix, and SWOT. \\
\bottomrule
\end{tabular}
\end{adjustbox}
\end{table}

In sum, the cross-case results indicate that governance determines \emph{where} evidence must land to become binding and \emph{who} has the authority to act on it. The case anchors further clarify that ports can operationalise LL work through different meso-level entry points (e.g.\ an Energy Community pilot versus a Green Coordinator function), while drawing on repeatable micro-level methods and tools within the same overarching macro--meso--micro LL logic.

\section{Discussion and future work}
\label{sec:discussion}

This study advances the port innovation literature by making explicit the linkage between governance arrangements and the design and institutionalisation of LLs. While LLs are widely invoked as vehicles for co-creation and experimentation \cite{Engels2019,ENoLL2025}, existing work has tended to treat governance primarily as context \cite{ENoLL2025,Madon2025} rather than as a set of actionable conditions that shape what LLs can achieve. The contribution here is twofold. First, the study specifies a governance--LL fit framework that connects ownership and decision rights to the four LL pillars (co-creation, real-life setting, iterative learning, and institutional embedding (multi-level decision-making)) and operationalises this connection through transparent indicators and a 0--2 scoring scheme. Second, it provides comparative, port-specific evidence showing how different governance logics create distinct institutional ``landing zones'' for LL outputs, thereby conditioning whether pilots remain episodic or translate into contract-grade or procedural change. To our knowledge, this is the first port study to integrate a governance typology with LL methodology and to demonstrate how that linkage influences the potential for innovation at both tenant and system levels.

\subsection{Positioning within and against the literature}

Empirical port studies have typically foregrounded operational efficiency, market structure, and ownership reforms (e.g.\ corporatisation, concessioning), treating innovation as a project-level attribute rather than an institutional process \citep{Heaver1995,NotteboomWinkelmans2001,BrooksCullinane2007,Verhoeven2010,DeLangen2004,NgPallis2010}. Recent work on sustainability and digitalisation in ports recognises new capabilities (green port strategies, port community systems, data platforms), yet often stops short of explaining how pilots become binding rules or routines \citep{Acciaro2014,Carlan2016,FruthTeuteberg2017,Heilig2017,LamNotteboom2014}. Our contribution complements this literature by specifying a governance--LL fit that makes the ``landing zones'' for evidence explicit (contracts, SOPs, procurement), helping explain why similar demonstrations yield different governance outcomes across port governance types.

The argument also extends public-sector innovation and collaborative governance scholarship by showing \emph{how} co-creation is converted into adoption under different port governance conditions. Across the two cases, LL outputs became durable only when they could be translated into concrete governance artefacts---for example, concession or lease annexes, procurement specifications, SOP updates, and coordination agreements. This moves beyond accounts that focus primarily on collaboration quality by identifying the practical ``landing zones'' through which evidence becomes binding.

Our results further suggest that the relevant mechanisms sit at the meso and macro layers of LL governance. At the meso level, ports can anchor LL work in a concrete entry point (e.g.\ PoA’s Energy Community; PoT’s Green Coordinator) that structures priorities and stakeholder roles. At the macro level, governance arrangements determine the authority, timing windows, and enforceability of the artefacts through which LL learning can be institutionalised. In short, the same micro-level LL methods (workshops, mapping, iterative testing) can produce different governance outcomes depending on where decision rights and codification instruments reside.

\subsection{Limitations and future work}
Several limitations temper the conclusions. The evidence reflects early-stage capacity-building and workshop artefacts rather than long-run outcome evaluations. The two cases were selected for analytic contrast; they are not statistically representative of all governance forms or regional contexts. While the scoring scheme improves transparency, it abstracts from political economy factors (for example, contested objectives, legacy contracts) that may influence the pace and direction of change.

Future research will deepen and broaden the empirical base along two lines. First, we will conduct and analyse more in-depth stakeholder interviews using the value-proposition reflection sheet and a semi-structured interview guide to validate and extend workshop-derived findings, strengthen the chain of evidence, and refine indicators. Second, we will recruit additional ports across governance types, including hybrid and joint-venture variants, to test the portability of the governance--LL fit framework, probe boundary conditions, and assess how interoperability standards support diffusion across institutional settings. Longitudinal follow-up will track whether artefacts proposed here (for example, concession/lease annexes or SOP updates) are adopted, under what timing and negotiation conditions, and with what operational and system-level effects. A further route to strengthen robustness is to expand within-port triangulation by analysing multiple pilots within the same port (e.g.\ several meso-level initiatives) and to conduct more controlled cross-port comparisons by examining the same type of case anchor (e.g.\ Energy Communities) across ports operating under different governance models.

\section*{Declaration of generative AI and AI-assisted technologies in the manuscript preparation process}
During the preparation of this work the author(s) used ChatGPT in order to improve clarity and readability of the manuscript, including assistance with language editing and phrasing. After using this tool/service, the author(s) reviewed and edited the content as needed and take(s) full responsibility for the content of the published article.

\renewcommand\refname{References}
\Urlmuskip=0mu plus 1mu\relax

\bibliographystyle{IEEEtran}
\bibliography{Ref}

@article{Monios2024,
   author = {Jason Monios and Gordon Wilmsmeier and Gustavo Andr{\'e}s Mart{\'i}nez Tello and Lara Pomaska},
   doi = {10.1016/j.jtrangeo.2024.103988},
   issn = {09666923},
   journal = {Journal of Transport Geography},
   month = {10},
   publisher = {Elsevier Ltd},
   title = {A new conception of port governance under climate change},
   volume = {120},
   year = {2024}
}

@article{Munim2020,
   author = {Ziaul Haque Munim and Henrik Sornn-Friese and Mariia Dushenko},
   doi = {10.1016/j.jclepro.2020.122156},
   issn = {09596526},
   journal = {Journal of Cleaner Production},
   month = {9},
   publisher = {Elsevier Ltd},
   title = {Identifying the appropriate governance model for green port management: Applying Analytic Network Process and Best-Worst methods to ports in the Indian Ocean Rim},
   volume = {268},
   year = {2020}
}

@article{Sornn-Friese2024,
   author = {Henrik Sornn-Friese},
   isbn = {9788793262164},
   pages = {55-69},
   publisher = {CBS Maritime},
   title = {Content, Structure and Governance of Transactions: A Business Model},
   url = {https://research.cbs.dk/en/publications/content-structure-and-governance-of-transactions-a-business-model},
   year = {2024}
}

@techReport{ENoLL2025,
   author = {ENoLL},
   title = {Collaborative Governance Framework Foundations of Living Labs \& Capacity Building. A Report submitted to POTENT-X},
   year = {2025}
}

@techReport{WorldBank2007,
   author = {{World Bank}},
   doi = {10.1596/978-0-8213-6607-3},
   title = {Port Reform Toolkit},
   url = {https://ppp.worldbank.org/public-private-partnership/library/port-reform-toolkit-ppiaf-world-bank-2nd-edition},
   year = {2007}
}

@book{Notteboom2021,
   author = {Theo Notteboom and Athanasios Pallis and Jean-Paul Rodrigue},
   city = {London},
   doi = {10.4324/9780429318184},
   isbn = {9780429318184},
   month = {12},
   publisher = {Routledge},
   title = {Port Economics, Management and Policy},
   year = {2021}
}

@article{Engels2019,
   author = {Franziska Engels and Alexander Wentland and Sebastian M. Pfotenhauer},
   doi = {10.1016/j.respol.2019.103826},
   issn = {00487333},
   issue = {9},
   journal = {Research Policy},
   month = {11},
   publisher = {Elsevier B.V.},
   title = {Testing future societies? Developing a framework for test beds and living labs as instruments of innovation governance},
   volume = {48},
   year = {2019}
}

@techReport{Madon2025,
   author = {B{\'e}n{\'e}dicte Madon and Antonio Torralba and Anas Alamoush and Lars J{\o}ker and Palle Skyum and Frans Libertson},
   institution = {University of Serville},
   month = {7},
   title = {Ports as Energy Transition Hubs-Collaborative Governance Framework},
   url = {https://potent-x.eu/documents/},
   year = {2025}
}

@article{Heaver1995,
  author    = {Trevor D. Heaver},
  title     = {The Implications of Increased Competition for Port Policy and Management},
  journal   = {Maritime Policy \& Management},
  year      = {1995},
  volume    = {22},
  number    = {2},
  pages     = {125--133},
  doi       = {10.1080/03088839500000039}
}

@article{NotteboomWinkelmans2001,
  author    = {Theo E. Notteboom and Willy Winkelmans},
  title     = {Structural Changes in Logistics: How Will Port Authorities Face the Challenge?},
  journal   = {Maritime Policy \& Management},
  year      = {2001},
  volume    = {28},
  number    = {1},
  pages     = {71--89},
  doi       = {10.1080/03088830119192}
}

@book{BrooksCullinane2007,
  editor    = {Mary R. Brooks and Kevin Cullinane},
  title     = {Devolution, Port Governance and Port Performance},
  publisher = {Elsevier/JAI Press},
  address   = {Oxford},
  year      = {2007},
  isbn      = {9780762314490}
}

@article{Verhoeven2010,
  author    = {Patrick Verhoeven},
  title     = {A Review of Port Authority Functions: Towards a Renaissance?},
  journal   = {European Journal of Transport and Infrastructure Research},
  year      = {2010},
  volume    = {10},
  number    = {4},
  pages     = {408--428}
}

@article{DeLangen2004,
  author    = {Peter W. de Langen},
  title     = {Governance in Seaport Clusters},
  journal   = {Maritime Economics \& Logistics},
  year      = {2004},
  volume    = {6},
  number    = {2},
  pages     = {141--156},
  doi       = {10.1057/palgrave.mel.9100108}
}

@article{NgPallis2010,
  author    = {Adolf K. Y. Ng and Athanasios A. Pallis},
  title     = {Port Governance Reforms in Diversified Institutional Frameworks: Generic Solutions, Implementation Asymmetries},
  journal   = {Transport Reviews},
  year      = {2010},
  volume    = {30},
  number    = {4},
  pages     = {395--415},
  doi       = {10.1080/01441640903230912}
}

@article{Acciaro2014,
  author    = {Michele Acciaro and H{\'e}l{\`e}ne Vanelslander and Christa Sys and Theo Notteboom and Claude Ferrari and Antonio Roumboutsos},
  title     = {Environmental Sustainability in Seaports: A Framework for Successful Innovation},
  journal   = {Maritime Policy \& Management},
  year      = {2014},
  volume    = {41},
  number    = {5},
  pages     = {480--500},
  doi       = {10.1080/03088839.2014.932926}
}

@article{LamNotteboom2014,
  author    = {J. S. L. Lam and Theo Notteboom},
  title     = {The Greening of Ports: A Comparison of Port Management Tools Used by Leading Ports in Asia and Europe},
  journal   = {Transport Reviews},
  year      = {2014},
  volume    = {34},
  number    = {2},
  pages     = {169--189},
  doi       = {10.1080/01441647.2014.891162}
}

@article{Carlan2016,
  author    = {Valentina Carlan and Christa Sys and H{\'e}l{\`e}ne Vanelslander},
  title     = {How Port Community Systems Can Contribute to Port Competitiveness: Developing a Cost--Benefit Framework},
  journal   = {Research in Transportation Business \& Management},
  year      = {2016},
  volume    = {19},
  pages     = {51--65},
  doi       = {10.1016/j.rtbm.2016.03.009}
}

@article{FruthTeuteberg2017,
  author    = {Malte Fruth and Frank Teuteberg},
  title     = {Digitization in Maritime Logistics: What Is There and What Is Missing?},
  journal   = {Cogent Business \& Management},
  year      = {2017},
  volume    = {4},
  number    = {1},
  pages     = {1411066},
  doi       = {10.1080/23311975.2017.1411066}
}

@inproceedings{Heilig2017,
  author    = {Lars Heilig and Stefan Schwarze and Sebastian Vo{\ss}},
  title     = {An Investigation of Digital Transformation in the Port Industry},
  booktitle = {Proceedings of the 50th Hawaii International Conference on System Sciences (HICSS)},
  year      = {2017},
  pages     = {1341--1350},
  doi       = {10.24251/HICSS.2017.163}
}

@article{PolydoropoulouEtAl2025PortsLLResilience,
  author  = {Polydoropoulou, Amalia and Bouhouras, Efstathios and Papaioannou, Georgios and Karakikes, Ioannis},
  title   = {Living labs for the resilience of ports against climate change disruptions},
  journal = {Ocean \& Coastal Management},
  year    = {2025},
  volume  = {261},
  pages   = {107528},
  doi     = {10.1016/j.ocecoaman.2024.107528},
  url     = {https://www.sciencedirect.com/science/article/abs/pii/S0964569124005131}
}

@article{GerlitzMeyerHenesey2024SMSPsLivingLabs,
  author  = {Gerlitz, Laima and Meyer, Christopher and Henesey, Lawrence},
  title   = {Sourcing Sustainability Transition in Small and Medium-Sized Ports of the Baltic Sea Region: A Case of Sustainable Futuring with Living Labs},
  journal = {Sustainability},
  year    = {2024},
  volume  = {16},
  number  = {11},
  pages   = {4667},
  doi     = {10.3390/su16114667},
  url     = {https://www.mdpi.com/2071-1050/16/11/4667}
}

@article{CatanaMerloPerboli2025GoverningDigitalTransformationPorts,
  author  = {Catana, Eusebiu and Merlo, Francesca and Perboli, Guido},
  title   = {Governing digital transformation in ports: Policy learning from the {5G-LOGINNOV} project},
  journal = {Open Research Europe},
  year    = {2025},
  volume  = {5},
  pages   = {362},
  doi     = {10.12688/openreseurope.21641.1},
  url     = {https://open-research-europe.ec.europa.eu/articles/5-362}
}

@online{MPA2026LivingLab,
  author  = {{Maritime and Port Authority of Singapore}},
  title   = {MPA Living Lab},
  year    = {2026},
  url     = {https://www.mpa.gov.sg/maritime-singapore/innovation-and-r-d/milestones-key-projects/mpa-living-lab},
  note    = {Accessed 7 January 2026}
}

@online{WPSP2023PIERLivingLab,
  author  = {{World Port Sustainability Program}},
  title   = {Halifax Port Authority -- The PIER living lab},
  year    = {2023},
  url     = {https://sustainableworldports.org/project/halifax-port-authority-the-pier-living-lab/},
  note    = {Accessed 7 January 2026}
}

@misc{ENoLL2025Origins,
  author       = {{European Network of Living Labs} and Schuurman, Dimitri and DeLosR{\'\i}os-White, Marta Irene and Desole, Martina},
  title        = {Living Lab origins, developments, and future perspectives},
  year         = {2025},
  publisher    = {Zenodo},
  doi          = {10.5281/zenodo.14764597},
  url          = {https://doi.org/10.5281/zenodo.14764597}
}

@techreport{CavalliniEtAl2016QuadrupleHelix,
  author       = {Cavallini, Simona and Soldi, Rossella and Friedl, Julia and Volpe, Margherita},
  title        = {Using the Quadruple Helix Approach to Accelerate the Transfer of Research and Innovation Results to Regional Growth},
  institution  = {European Union / European Committee of the Regions},
  year         = {2016},
  doi          = {10.2863/408040},
  isbn         = {978-92-895-0890-2},
  url          = {https://op.europa.eu/en/publication-detail/-/publication/6e54c161-36a9-11e6-a825-01aa75ed71a1}
}

@phdthesis{Schuurman2015PhD,
  author       = {Schuurman, Dimitri},
  title        = {Bridging the gap between Open and User Innovation? Exploring the value of Living Labs as a means to structure user contribution and manage distributed innovation},
  school       = {Ghent University and Vrije Universiteit Brussel},
  year         = {2015},
  url          = {http://hdl.handle.net/1854/LU-5931264}
}

@article{FauthDeMoortelSchuurman2024Orchestrators,
  author       = {Fauth, Janin and De Moortel, Kevin and Schuurman, Dimitri},
  title        = {Living labs as orchestrators in the regional innovation ecosystem: a conceptual framework},
  journal      = {Journal of Responsible Innovation},
  year         = {2024},
  volume       = {11},
  number       = {1},
  pages        = {2414505},
  doi          = {10.1080/23299460.2024.2414505},
  url          = {https://doi.org/10.1080/23299460.2024.2414505}
}

\newpage
\appendix 
\section*{Appendix}

\clearpage
\section{Indicator codebook (0--2 scale)}
\label{app:codebook}

\addcontentsline{toc}{section}{Appendix A: Indicator codebook (0--2 scale)}

\subsection*{Dimension 1 --- Stakeholder dynamics}
\noindent\textbf{Indicators:} representation across the Quadruple Helix (public sector, industry/tenants, academia, civil society/users); engagement frequency and continuity; role clarity and communication channels; conflict-resolution processes.

\noindent\textbf{Scoring rubric:}
\begin{itemize}
  \item \textbf{0:} Narrow or ad hoc engagement; unclear roles; irregular interaction.
  \item \textbf{1:} Targeted but consistent engagement; roles partly specified; facilitated sessions.
  \item \textbf{2:} Broad and continuous engagement; explicit role maps; standing mechanisms for resolution.
\end{itemize}

\subsection*{Dimension 2 --- Decision authority and embedding}
\noindent\textbf{Indicators:} decision loci and budget authority; formal landing zones for codifying results (concessions/leases, procurement, SOPs, regulations); identifiable timing windows (renewals, CAPEX, budget cycles); traceable evidence pathways.

\noindent\textbf{Scoring rubric:}
\begin{itemize}
  \item \textbf{0:} Opaque authority; no clear channels; timing unpredictable.
  \item \textbf{1:} Identifiable internal channels; external codification ad hoc; partial timing alignment.
  \item \textbf{2:} Multiple, well-defined channels; explicit timing windows; traceable evidence pathways.
\end{itemize}

\subsection*{Dimension 3 --- Innovation scope}
\noindent\textbf{Indicators:} balance of tenant-level and system-level pilots; energy--digital coupling (metering, data models, controls, interoperability); operational scale in real-life settings; external linkages (DSO/TSO, municipal energy planning, logistics interfaces).

\noindent\textbf{Scoring rubric:}
\begin{itemize}
  \item \textbf{0:} Isolated, small pilots; weak data or energy linkages.
  \item \textbf{1:} Solid internal pilots; selective external links; moderate scale.
  \item \textbf{2:} Balanced portfolio; strong coupling; system-level integration possibilities.
\end{itemize}

\subsection*{Dimension 4 --- Institutional uptake mechanisms}
\noindent\textbf{Indicators:} instruments available (contracts, concessions/leases, procurement clauses, SOPs, MOUs); specification quality (measurable KPIs, data-sharing terms, monitoring); enforcement and review processes; replicability through templates or patterns.

\noindent\textbf{Scoring rubric:}
\begin{itemize}
  \item \textbf{0:} No reusable mechanisms; informal uptake only.
  \item \textbf{1:} Internal procedures and procurements; limited external enforceability.
  \item \textbf{2:} Contract-grade instruments with KPIs and monitoring; templates for diffusion.
\end{itemize}


\clearpage
\section{Cross-case matrix (governance--LL fit)}
\label{app:matrix}

\begin{table}[h!]
\centering
\small
\begin{tabular}{p{2cm} p{3.5cm} >{\centering\arraybackslash}p{1.2cm} p{3.5cm} >{\centering\arraybackslash}p{1.2cm} p{3.5cm}}
\hline
Dimension & Indicators (summary) & PoA (R1/R2) & Evidence anchors & PoT (R1/R2) & Evidence anchors \\
\hline
Stakeholder dynamics &
Breadth across Quadruple Helix; intensity and cadence; role clarity; resolution mechanisms &
2/1 &
R1: Multi-actor canvases; recurring joint sessions; explicit role mapping. R2: Strong stakeholder network; developing collaboration mechanisms. &
1/0 &
R1: Targeted participation; project-specific involvement; facilitation notes. R2: Narrow interaction; self-focused targets. \\
Decision authority and embedding &
Decision logic; availability of landing zones (concessions, procurement, SOPs); timing windows; traceability &
2/1 &
R1: Concession and lease pathways; procurement playbooks; decision timelines. R2: Clear workflows; process bottlenecks. &
1/1 &
R1: Internal SOPs; municipal coordination; limited external codification channels. R2: Clear authority; limited financial instruments. \\
Innovation scope &
Tenant versus system pilots; energy--digital coupling; operational scale; external linkages &
2/2 &
R1: Tenant pilots and system projects; data platform linkage. R2: Significant innovation scale; systemic innovation landscape. &
1/1 &
R1: Internal pilots; selective DSO or municipal interfaces. R2: Internal scope; moderate ambitions. \\
Institutional uptake mechanisms &
Instruments; specification quality; enforcement and review; replicability &
2/2 &
R1: Concession annexes; KPI clauses; port-wide procedures. R2: Clear external interfaces. &
1/1 &
R1: SOP updates; framework procurements; memoranda of understanding. R2: Internal procedures. \\
\hline
\end{tabular}
\caption{Cross-case comparison of governance--LL fit with two independent coder scores (R1/R2). Scores: 0 = low/absent; 1 = moderate/mixed; 2 = high/strong.}
\end{table}

\end{document}